\documentstyle[12pt]{article}
\newcommand{\be}{\begin{equation}}
\newcommand{\ee}{\end{equation}}
\begin{document}
\title{Nonlinear Impurity in a Lattice: Dispersion Effects}
\author{{\bf M. I.  Molina$^{\dagger}$}
\vspace{1 cm}
\and
\and
Facultad de Ciencias, Departamento de F\'{\i}sica, Universidad de Chile\\
Casilla 653, Las Palmeras 3425, Santiago, Chile.\\
mmolina@abello.dic.uchile.cl}
\date{}
\maketitle
\baselineskip 18 pt
\begin{center}
{\bf Abstract}
\end{center}
\noindent
We examine the bound state(s) associated with a single cubic nonlinear impurity, in a
one-dimensional tight-binding lattice, where hopping to first--and--second nearest neighbors
is allowed. The model is solved in closed form {\em v\`{\i}a} the use of the appropriate lattice Green function  
and a phase diagram is obtained showing the number of bound states as a function of
nonlinearity strength and the ratio of second to first nearest--neighbor hopping parameters.
Surprisingly, a finite amount of hopping to second nearest neighbors helps the formation
of a bound state at smaller (even vanishingly small) nonlinearity values.
As a consequence, the selftrapping transition can also be tuned to occur at relatively
small nonlinearity strength, by this increase in the lattice dispersion.
\vspace{1cm}

\noindent
$^{\dagger}$email: mmolina@abello.dic.uchile.cl
\vspace{2 cm}

\noindent
PACS number(s):\ \ 71.55.-i

\newpage

The effects of impurities on the transport properties of materials continues to be an
interesting subject. When the concentration of impurities inside a material is finite,
we speak of a disordered system. In one dimensional systems, this disorder gives rise
to the well-known phenomenon of Anderson localization, where all states acquire a finite
localization length. This precludes any amount of transport in the system. Economou
and coworkers\cite{economou1} have argued that there exist a correspondence between
the property that, in dimensions smaller or equal to two, all states of a disordered
system are localized, and the property that an isolated impurity can always trap a particle.
More recently, the problem of {\em nonlinear} impurities have received considerable attention.
In a condensed matter context, they appear in strongly coupled electron--vibration systems,
when the vibrational degrees of freedom have the ability of adapting quickly to the presence
of the electron, giving rise to polaronic behavior\cite{cmt}. Nonlinear impurities appear also in
other fields, such as nonlinear optics. For instance, an array of linear waveguides containing
a single or several nonlinear, Kerr-like guides. The transversal dynamics describing the
energy exchange among waveguides is formally identical to the dynamics of an excitation
propagating in a linear tight-binding lattice in the presence of one or more nonlinear
impurities\cite{tdm}. 

In systems where an electron (or excitation) is propagating while strongly interacting with
vibrational degrees of freedom, an ``effective'' nonlinear evolution equation for the electron
can be obtained, under the assumption that the vibrations adapt instantly to the presence of the
electron\cite{cmt}. This equation, known as the Discrete Nonlinear Schr\"{o}dinger (DNLS) equation,
has the form:
\be
i\ \left( {d C_{{\bf n}}\over{d t}} \right) = V \sum_{\bf n.n.}  C_{{\bf m}} - \chi_{{\bf n}}
| C_{{\bf n}} |^{2}\ C_{{\bf n}} \hspace{1cm}(\hbar \equiv 1),\label{eq:0}
\ee
where $C_{\bf n}$ is the probability amplitude of finding the electron on lattice site ${\bf n}$ at time $t$,
$V$ is the nearest-neighbor hopping parameter and $\chi_{{\bf n}}$ is the nonlinearity parameter at site
${\bf n}$ proportional to the square of the electron-vibration coupling. In the conventional DNLS
equation, the sum in (\ref{eq:0}) is restricted to nearest-neighbors (n.n.).

For the case of a single nonlinear impurity in a onedimensional lattice ($\chi_{n} = \delta_{n,0}\ \chi$),
we have obtained in a previous work\cite{mt_prb} its bound state
analytically, {\em v\`{\i}a} lattice Green functions, and have shown that a bound state is possible provided
$|\chi/2 V| > 1$. This result deviated markedly from the well-known linear impurity case where, a bound state
exists for any impurity strength.  The extension to an impurity of arbitrary nonlinear exponent
$\alpha$ (i.e., $| C_{{\bf n}} |^{\alpha}$ instead of $| C_{{\bf n}} |^{2}$ in (\ref{eq:0})),
revealed\cite{mth_pre,saopaulo} that
for $\alpha < 2$ there is always a bound state for any finite $\chi/V$. At $\alpha = 2$ (the standard
DNLS case) there is one bound state for $|\chi/2V| > 1$, while for $\alpha > 2$ there is a critical curve
in $\chi$--$\alpha$ space, below which there is no bound state, while above it, there are two bound states.
On the critical curve, there is a single bound state.  

When this nonlinear impurity is embedded in a square lattice\cite{square}, the
$\chi$--$\alpha$ bound state phase diagram shows a single curve separating two
regimes. Below the curve, there are no bound states; on the curve there is a single
bound state, while above the curve there are two bound states. One of these become
more localized upon increment of the nonlinearity parameter while the other becomes
more delocalized. Bound states for single nonlinear impurities have also been computed
for other systems, including a Cayley tree\cite{cayley}, a triangular lattice\cite{busta}
and a cubic lattice\cite{hui, busta}.

In all the studies above, only dispersion to first nearest--neighbors has been considered.
For systems with long-range dispersion, a continuum-like approximation that employs
a nonlocal nonlinear Schr\"{o}dinger equation has been proposed\cite{gaididei}. For
the case of a discrete system with a hopping parameter of the form $V_{n m} = V/| n - m|^{s}$,
another study\cite{rasmussen} followed a variational approach, based on a
plausible {\em ansatz} for the localized state. Among other things, they found that
there is a critical $s_{cr}$ such that all dispersive interactions decreasing faster
than $r^{-s{cr}}$ lead to similar qualitative behavior as the DNLS with only nearest
neighbor transfer. In this work we solve in closed form {\em v\`{\i}a} Green functions,
the case of a single DNLS impurity in a tight-binding lattice, including hopping to
first and second nearest--neighbors.  Since the phenomenon of selftrapping is the result
of the quantum struggle between the tendency to spread (dispersion) and the tendency to
localize (nonlinearity), one might surmise that any increase in dispersion will have the
simple effect of increasing the nonlinearity needed to selftrap. However as we will see,
this is not necessarily the case and a small increment in dispersion can actually favor
the formation of a bound state at smaller nonlinearity strength.

\section{Onedimensional lattice with dispersion}

Let us consider the problem of determining the existence of bound states and
dynamic selftrapping characteristics for an electron (or an excitation)
moving on a onedimensional dispersive lattice with hopping up to second
nearest-neighbors, which contains a single DNLS
impurity at the origin $n = 0$. The DNLS equation (\ref{eq:0}) reduces to
\be
i\ \left( {d C_{n}\over{d t}} \right) = V_{1} ( C_{n+1} + C_{n-1} ) + V_{2} ( C_{n+2} + C_{n-2}) -
\chi| C_{0} |^{2}\ C_{0}\ \delta_{n,0}.\ \ \label{eq:dnls}
\ee
For stationary states, one puts $C_{n}(t) = \exp(-i E t)\ \phi_{n}$, obtaining
\be
E\ \phi_{n} = V_{1} (\phi_{n+1} + \phi_{n-1}) + V_{2}( \phi_{n+2}+\phi_{n-2}) - \chi\ |\phi_{0}|^{2}\ \phi_{0}\ \delta_{n 0}.
\ee

The Hamiltonian that gives rise to (\ref{eq:dnls}) is
\begin{equation}
{\tilde{H}}={\tilde{H}}_0+{\tilde{H}}_1,
\end{equation}
where 
\begin{equation}
{\tilde{H}}_0 = V_{1}\; \sum_{n} ( |n\rangle\langle n+1| + |n+1\rangle\langle n|\; ) +
	        V_{2}\; \sum_{n} ( |n\rangle\langle n+2| + |n+2\rangle\langle n|\; )\label{eq:H0}
\end{equation}
and 
\begin{equation}
{\tilde{H}}_1 = -\chi |\phi_{0}|^{2}\; |0\rangle\langle 0|.
\end{equation}
The $\{|n\rangle\}$ represent Wannier electronic states and 
$V_{1}$($V_{2}$) is the nearest (next-to-nearest) neighbor transfer matrix element.
In the absence of impurity, the energy band is given by
\be
E(k) = 2\ V_{1}\ \cos(k) + 2\ V_{2} \cos(2 k).\label{eq:ek}
\ee
A simple analysis shows that, for positive $V_{2}/V_{1}$, the upper and lower band edges obey
\be
E_{max} = 2 (V_{1} + V_{2})\label{eq:emax}
\ee
\be
E_{min} = \left\{ \begin{array}{ll}
	-2(V_{1} - V_{2}) 	& \mbox{$V_{2}/V_{1} < 1/4$}\\
	-(V_{1}^{2}/4 V_{2}) - 2 V_{2}	& \mbox{$V_{2}/V_{1} > 1/4$}
		\end{array}
			\right.\label{eq:emin}
\ee
As a result, while the upper edge always increase linearly with $V_{2}$, the lower edge
first decreases (in magnitude),
reaching a minimum value of $2 (1-(1/\sqrt{8})) V_{1} \approx 1.29 V_{1}$ at $V_{2} = (1/\sqrt{8}) V_{1}$.
Afterwards, the lower edge increases in magnitude with $V_{2}$. At large $V_{2}$, this increase will be
almost linear. These features will of importance in
the next section, where we determine the position of the impurity bound state(s).

The formalism of lattice Green functions is particularly suitable for the
problem of determining the bound state(s). First, we normalize all energies
by $V_{1}$ and define: $z\equiv E/V_{1}$, $H\equiv {\tilde{H}}/V_{1}$,
$\gamma \equiv \chi /V_{1}$ and $\delta \equiv V_{2}/V_{1}$.
The dimensionless lattice Green function $%
SG = 1/(z-H)$ can be formally expanded as\cite{economou2} 
\begin{equation}
G=G^{(0)}+G^{(0)}H_1G^{(0)}+G^{(0)}H_1G^{(0)}H_1G^{(0)}+... \label{eq:series}
\end{equation}
where $G^{(0)}$ is the unperturbed ($\gamma =0$) Green function and
$H_1 = -\gamma |\phi_{0}|^{2} \; |0\!><~\!\!0|$. The sum in
Eq.(\ref{eq:series}) can be carried out exactly to yield 
\begin{equation}
G_{m n} = G_{m n}^{(0)} - {\frac{\gamma |\phi_{0}|^{2} \;
G_{m 0}^{(0)}\; G_{0 n}^{(0)}}{{%
1 + \gamma |\phi_{0}|^{2} \; G_{0 0}^{(0)}}}}.  \label{eq:gtotal}
\end{equation}
where $G_{m n} = \langle m| G |n \rangle$ and
\be
G_{m n}^{(0)}(z) = {1\over{2 \pi}}\int_{-\pi}^{\pi}\ d\phi\ {\exp[ i \phi (m-n)\ ]\over{
[\ z - \cos(\phi) - \delta\ \cos(2 \phi)}\ ]}.\label{eq:G0}
\ee
whose evaluation in closed form is done in the {\em Appendix}.

\subsection{Bound States}

The energy of the bound state(s), $z_b$ is obtained
from the poles of $G_{m n} $, {\em i.e.}, by solving $1 =
-\gamma |\phi_{0}^{(b)}|^{2} \; G_{0 0}^{(0)}$.
The bound state amplitudes $\phi_{n}^{(b)}$ are obtained from the
residues of $G_{m n}(z)$ at $z = z_b$. In particular,  
\be
|\phi_{0}^{(b)}|^2 = \mbox{Res}\{G_{0 0}(z)\}_{z=z_b} =
- {{G_{0 0}^{(0)}}^{2} (z_{b}) \over{G_{0 0}^{'(0) } (z_{b})}}
\label{eq:c2}
\ee
Inserting this in the bound state energy equation leads to
\be
{1\over{\gamma}} = {{G_{0 0}^{(0)}}^{3} (z_{b})\over{
G_{0 0 }^{'(0)} (z_{b})\; }}.\label{eq:zb}
\ee
Now, as the reader can easily verify using (\ref{eq:zb}) and (\ref{eq:G0}), changing both, the sign of $\chi/V_{1}$ and the sign of $V_{2}/V_{1}$
has the effect of changing the sign of the corresponding eigenvalue. Thus, for a complete parameter space
examination of all the possible eigenvalues, we only need to consider a fixed sign for $V_{2}/V_{1}$ (say, positive)
and the two possible signs of $\chi/V_{1}$.

Figure 1 shows the left- and right-hand side of (\ref{eq:zb}) for several (positive) values of
$\delta$. When $\gamma > 0$ (Fig. 1a) and as $\delta$ increases from zero, the RHS of
(\ref{eq:zb}) moves towards the origin and increases its height, until $\delta$ reaches $1/4$,
where the height diverges. Further increase in $\delta$ decreases the height of the curves, but
they continue to approach the origin until $\delta = 1/\sqrt{8}$. Afterwards, the curves move away
from the origin while their heights continue to decrease. For the case of $\gamma < 0$ (Fig. 1b),
the situation is quite different: For a given $\delta$ value, there is a minimum $|\gamma_{a}|$ at which there
is a bound state.
Further increase in $|\gamma|$ creates two bound states, one of which will ultimately disappear
at a further finite $|\gamma_{b}|$ value, leaving only a single bound state.

Figure 2 displays a phase diagram in nonlinearity--dispersion space showing
the number of bound state(s). For positive nonlinearity, the critical curve separating
the region with no bound states from the region with one bound state, decreases with
increasing $\delta$ and reaches zero at $\delta = 1/4$. Afterwards, it increases
monotonically with further dispersion. Thus, there is a finite dispersion interval,
$0 < \delta < 1/4$ where, contrary to what might be expected, an increase in
dispersion actually {\em reduces} the nonlinearity needed to create a bound state.
This can also be seen in Fig. 3a, which shows the bound state energy as a function
of (positive) nonlinearity, for several values of dispersion $\delta$.
This reduction in nonlinearity needed to sustain a bound state is, of course
due to the reduction in the width of the negative portion of the band with a small
positive dispersion, and thus, it has a {\em linear} origin.
In the negative nonlinearity sector, we have in Fig.2 two critical curves
separating regions with no bound states, two bound states and one bound state. Here,
an increase in dispersion, causes a corresponding increase in the minimum nonlinearity
needed for the creation of a bound state(s).
This is correlated with the fact that the width of the positive portion of the
band always increases linearly with $\delta$. Figure 3b, shows the bound state energy as a function
of nonlinearity, for negative $\gamma$. Here, for a given value of
$\delta$, there exists a critical nonlinearity value $\gamma_{a}$ for which there is a bound state, with
energy outside the band. Further increase in nonlinearity creates two bound states, one of which
increases its energy monotonically with nonlinearity while the other state decreases its energy
towards the band, reaching it at a finite nonlinearity value $\gamma_{b}$. As Fig. 4 shows, in the regime
with two bound states, as the magnitude of the nonlinearity is increased, one of the states
becomes more localized on the impurity site, while the other becomes more delocalized,
ultimately disappearing into the continuum at a finite nonlinearity value.

\subsection{Transmission across the impurity}

Inside the band, $z_{min}(\delta) < z < z_{max}(\delta)$ (see (\ref{eq:emax}) and (\ref{eq:emin})),
all states are extended and given by\cite{economou2}
\be
|\psi\rangle = |k\rangle + G^{(0)+}\ T^{+}(z)\ |k\rangle,
\ee
where $|k\rangle$ is a plane wave and the scattering--matrix $T$ is
\be
T = H_{1} + H_{1}\ G^{(0)}\ H_{1} + H_{1}\ G^{(0)}\ H_{1}\ G^{(0)}\ H_{1} + ...
\ee
This can be summed exactly, to yield
\be
T = {-\gamma |C_{0}|^{2}|0\rangle \langle 0|\over{1 + \gamma |C_{0}|^{2}\ G_{0 0}^{(0)}}}.
\ee
With this, we can compute the scattering amplitude at any site $n$. In particular, the
scattering amplitude at the {\em impurity site} $C_{0} = \langle 0|\Psi\rangle$ is
\be
C_{0} = \langle 0|k\rangle -
{\gamma |C_{0}|^{2}\ \langle 0|G^{(0)+}(z)|0\rangle\langle0|k\rangle\over{1 + \gamma |C_{0}|^{2}\ G_{0 0}^{(0)}}}.\label{eq:c0}
\ee
The transmission coefficient is then given by the probability at the impurity site,
$t = |C_{0}|^{2}$. From (\ref{eq:c0}), we have
\be
t = {1\over{| 1 + \gamma\ t\ G_{0 0}^{(0)} |^{2}}} = {1\over{1 + \gamma^{2}\ t^{2}\ \mbox{Im}[G_{0 0}^{0)}(z)]^{2}}}
\ee
which leads to the cubic equation: $\gamma^{2} t^{3} \mbox{Im}[G_{0 0}^{0)}(z)]^{2} + t -1 = 0$.
This is invariant under $\gamma\rightarrow -\gamma$, implying that the transmission does not
depend on the sign of the nonlinearity parameter. The physical solution for $t$ is
\be
t = \frac{-2\ 6^{\frac{1}{3}} + {\left( 18\,\gamma \mbox{Im}[G_{0 0}^{0)}(z)] + 
        2\,{\sqrt{3}}\,{\sqrt{4 + 27\,\gamma^{2} \mbox{Im}[G_{0 0}^{0)}(z)]^{2}}} \right) }^{\frac{2}{3}}}
    {2\ 3^{\frac{2}{3}}\,\gamma\ \mbox{Im}[G_{0 0}^{0)}(z)]\,{\left( 9\,\gamma \mbox{Im}[G_{0 0}^{0)}(z)] + 
        {\sqrt{3}}\,{\sqrt{4 + 27\,\gamma^{2} \mbox{Im}[G_{0 0}^{0)}(z)]^{2}}} \right) }^{\frac{1}{3}}}.
\ee
The specific form of Im[$G_{0 0}^{(0)}(z)$] in our case, is given in closed form in the {\em Appendix}.
Figure 5 shows several transmission curves as functions of the plane waves dimensionless energy $z$, for several
different ratios $\delta$. The most remarkable new feature is the appearance of an abrupt ``dip'' on the
transmission near the lower edge of the band at $\delta \sim 0.4$. As $\delta$ increases further,
the ``dip'' moves to the right and eventually (not shown) approaches the upper band edge and merges with it.
This ``dip'' is related to the creation of a secondary ``branch'' in Im[$G_{0 0}^{(0)}(z)$],
as shown in Fig. B of the {\em Appendix}.

\subsection{Dynamic Selftrapping}

We place the electron at the impurity site at $t=0$ and observe its time evolution,
according to Eq.(\ref{eq:dnls}). The observable of interest here is the long-time
average probability of finding the electron on the initial site after a relatively
long time $T$:
\be
P_{0} =
\lim_{T\rightarrow \infty} (1/T)\ \int_{0}^{T} | C_{\bf 0}(t) |^{2} dt, \hspace{1cm}
| C_{\bf 0}(0) | = 1.
\ee
By using the transformation $C_{n}\rightarrow (-1)^{n}\ C_{n}$, that leaves $|C_{n}|^{2}$
invariant, the reader can verify that changing $V_{2}\rightarrow -V_{2}$ and $\chi\rightarrow
-\chi$, transforms (\ref{eq:dnls}) into its complex conjugate. Thus, for our completely
localized real initial condition $C_{0}(0) = 1$, the observable $P_{0}$ is invariant under the
transformation. Therefore, for a complete parameter study, it is enough to consider a positive $V_{2}$ (or $\delta$)
and the two possible signs for the anharmonicity $\chi$ (or $\gamma$).

We use a fourth-order Runge-Kutta numerical scheme, whose accuracy is monitored through
total probability conservation: $ 1 = \sum_{n} |C_{n}(t)|^{2}$.
To avoid undesired boundary effects, a self-expanding lattice is used\cite{mt_prl}.
Figure 6 shows $P_{0}$ as a function of (positive) nonlinearity parameter $\gamma$, for several
different dispersion $\delta$ values. As anticipated from the bound state results,
an increase of dispersion reduces the critical nonlinearity for the onset of selftrapping.
The minimum threshold occurs around $\delta \approx 0.3$. In the immediate vicinity of this value, the
selftrapping transition also seems to lose some sharpness.
Subsequent increase in $\delta$ increases the critical nonlinearity $\gamma_{c}$
again and restores sharpness to the $P_{0}$ curve.
At $\delta \approx 1$, the $P_{0}$ curve almost coincides with the $\delta = 0$ case.
Thereafter, $\gamma_{c}$ increases in an almost linear fashion with $\delta$.
For the case of a negative nonlinearity parameter (not shown),
the critical nonlinearity always increases monotonically with dispersion.
As explained above, this case corresponds also to $\gamma > 0, \delta < 0$.

\section{Discussion}

We have examined the conditions for the formation of a bound state at a nonlinear
impurity site, in a onedimensional linear lattice with hopping to first and second
nearest-neighbors. The formalism employed lattice Green functions, which have been
evaluated in closed form for our system. We found that this increment in dispersion
can actually favor the formation of a bound state at smaller nonlinearity strength.
As a consequence, the onset of dynamical selftrapping at the impurity site can also
be shifted to lower nonlinearity thresholds. This tuning effect could have some impact
on the design of completely nonlinear optical or solid nanostructures, where one is
interested in static (``selftrapped'') or mobile discrete nonlinear excitations
(``discrete solitons''). In the former, one is interested in achieving abrupt selftrapping
at relatively low nonlinearity (or input power) parameter values; in the latter, the
probability profile corresponding to an impurity bound state, could be useful as an initial
condition for the launch of a discrete soliton, which can carry energy across a nonlinear
fiber array from a given guide to another, distant one\cite{kivshar}.       

\section{Acknowledgments}

This work was supported in part by FONDECYT grant 1020139.

\newpage
\section{APPENDIX:\ THE LATTICE GREEN FUNCTION}

The dimensionless lattice Green function for our tight-binding chain including
dispersion up to second nearest-neighbors is
\be
G_{m n}^{(0)}(z) = {1\over{2 \pi}}\int_{-\pi}^{\pi}\ d\phi\ {\exp[ i \phi (m-n)\ ]\over{
[\ z - \cos(\phi) - \delta\ \cos(2 \phi)}\ ]}.\label{eq:G}
\ee
We follow the usual method\cite{economou1} of transforming (\ref{eq:G}) to an integral over a complex
variable $w$ along the unit circle. Thus, we have
\be
G_{m n}^{(0)}(z) = -{1\over{2 \pi i}} \oint {d w \ w^{|l-m|+1}\over{\delta w^{4} + w^{3} - w^{2} z + w + \delta}}
\ee

The roots of $\delta\  w^{4} + w^{3} - w^{2} z + w + \delta = 0$ are given by
\[
z_{1} = {-1 + \delta\ \gamma_{1}(z) - \sqrt{2} \sqrt{\gamma_{2}(z) - \delta\ \gamma_{1}(z)})\over{4 \delta}}
\]
\[
z_{2} = {-1 - \delta\ \gamma_{1}(z) + \sqrt{2}\sqrt{\gamma_{2}(z) + \delta\ \gamma_{1}(z)})\over{4 \delta}}
\]
\[
z_{3} = {-(1 + \delta\ \gamma_{1}(z) + \sqrt{2}\sqrt{\gamma_{2}(z) + \delta\ \gamma_{1}(z)})\over{4 \delta}}
\]
\[
z_{4} = {-1 + \delta\ \gamma_{1}(z) + \sqrt{2}\sqrt{\gamma_{2}(z) - \delta\ \gamma_{1}(z)})\over{4 \delta}}
\]
where $\gamma_{1}(z)\equiv \sqrt{8 + (1/\delta^{2})+(4 z/\delta)}$ and
$\gamma_{2}(z) \equiv 1-4 \delta^{2} + 2 \delta z$. Only the poles contained inside the unit circle
contribute to the integral. Thus,
\begin{eqnarray*}
|z_{1}| & < & 1 \Rightarrow\hspace{1cm} 2 (1 + \delta) < z < \infty\\
|z_{2}| & < & 1 \Rightarrow  \left\{ \begin{array}{ll}
		-\infty < z < \infty,\hspace{1cm} \delta<1/4\\
		-\infty < z < -1/(4 \delta) - 2 \delta \hspace{0.5cm}\mbox{and}\hspace{0.5cm} 2(\delta -1) < z < \infty,\hspace{0.5cm} \delta > 1/4
		\end{array}
		\right.\\
|z_{3}| & > & 1 \Rightarrow \mbox{no contribution}\\
|z_{4}| & < & 1 \Rightarrow \left\{ \begin{array}{ll}
			-\infty < z < -2 (1-\delta), \hspace{0.5cm} \delta<1/4\\
			-\infty < z < -1/(4 \delta) - 2 \delta, \hspace{0.5cm} \delta > 1/4
			\end{array}
			\right.		
\end{eqnarray*}

Using the above, we can express $G_{l m}^{(0)}(z)$ as
\begin{eqnarray}
\lefteqn{G_{l m}^{(0)}(z) = -{1\over{2 \pi \delta}} \oint {dw\ w^{|l - m| + 1}\over{\delta (w - z_{1})(w - z_{2})(w - z_{3})(w - z_{4})}}}\\
& &=  -{\Theta[ z-(2(1+\delta) ]\ z_{1}^{|l-m| + 1}\over{\delta (z_{1}-z_{2})(z_{1}-z_{3})(z_{1}-z_{4})}} +\nonumber\\
& & -{[\Theta((1/4)-\delta)+\Theta(\delta-(1/4))[\Theta(-1/(4 \delta-2 \delta - 2z))+\Theta(z-2(\delta-1))]]\ z_{2}^{|l-m| + 1}\over{\delta (z_{2}-z_{1})(z_{2}-z_{3})(z_{2}-z_{4})}} +\nonumber\\
& & -{[\Theta((1/4)-\delta) \Theta(-2(1-\delta)-z) + \Theta(\delta-(1/4)) \Theta(-1/(4 \delta)-2\delta-2 z)]\ z_{4}^{|l-m| + 1}\over{\delta (z_{4}-z_{1})(z_{4}-z_{2})(z_{4}-z_{3})}}\ \ \ \ \ 
\end{eqnarray}
After some tedious algebra, we find for the real and imaginary parts of the diagonal Green function:
\vspace{1cm}

\be
Re[G_{0 0}^{(0)}(z)] = \left\{ \begin{array}{ll}
2 \sqrt{2} \delta\ [ (\gamma_{1} \delta + \gamma_{2}) \sqrt{1-\delta\ (4 \delta - 2 z + \gamma_{1})} + \\
\hspace{1cm}(\gamma_{1} \delta - \gamma_{2})\sqrt{1 + \delta(-4 \delta + 2 z + \gamma_{1})}\ ]^{-1} & z > z_{max}(\delta)\\
\\~
- 2 \sqrt{2}\delta\ [ (\gamma_{1}\delta + \gamma_{2}) \sqrt{1 - \delta\ (4 \delta-2 z + \gamma_{1})} -\\
\hspace{1cm}(\gamma_{1}\delta - \gamma_{2}) \sqrt{1 + \delta\ (-4 \delta + 2 z + \gamma_{1})}\ ]^{-1} &  z < z_{min}(\delta)\\
\\~
0	& 	\mbox{otherwise}\ \ \ \ \ \ \ \ \ \ 
\end{array}
\right.
\ee
where,
\[
z_{min}(\delta) = \left\{ \begin{array}{ll}
-2( 1 - \delta)          & \delta < 1/4\\
-1/(4 \delta) - 2 \delta & \delta > 1/4
\end{array}
\right.
\]
and
\[
z_{max} = 2 ( 1 + \delta)
\]

\begin{eqnarray}
Im[G_{0 0}^{(0)}(z)] & = & {\Theta((1/4) - \delta) [\Theta(z - 2 (\delta-1)) - \Theta(z - 2 (1 + \delta))]
\over{
2\sqrt{1 - d_{1}(z)^{2}} |1 + 4 \delta\ d_{1}(z) |}} + \nonumber\\
                     &   & {\Theta(\delta - (1/4)) [\Theta(z + (1/(4 \delta)) + 2 \delta) - \Theta(z - 2 (1 + \delta)) ]
\over{
2\sqrt{1 - d_{1}(z)^{2}} | 1 + 4 \delta\ d_{1}(z) |}} + \nonumber\\
                     &   & {\Theta(\delta - (1/4))[\Theta(z + (1/(4 \delta)) + 2 \delta))-\Theta(z - 2 (\delta - 1)) ]
\over{
2 \sqrt{1 - d_{2}(z)^{2}} |1 + 4 \delta\ d_{2}(z) |}}\ \ \ \ \ \ \ \ \ \ 
\end{eqnarray}
where,
$d_{1}(z) \equiv (1/4\ \delta)( -1 + \sqrt{1 + 8\ \delta\ (\delta + (z/2))})$ and
$d_{2}(z) \equiv (1/4\ \delta)( -1 - \sqrt{1 + 8\ \delta\ (\delta + (z/2))})$.

\noindent
Figures A and B, show Re[$G_{0 0}^{(0)}(z)$] and Im[$G_{0 0}^{(0)}(z)$], respectively for
different values of dispersion $\delta$.

\newpage

\newpage

\centerline{{\bf Captions List}}
\vspace{2cm}

\noindent {\bf Fig.1 :}\ \ Left--and--right hand side of the eigenenergy equation (\ref{eq:zb}), for
several dispersion $\delta$ values, for positive (a) and negative(b) nonlinearity.
\vspace{0.4cm}

\noindent{\bf Fig.2 :}\ \  Bound state phase diagram in nonlinearity--dispersion space ($\gamma \equiv \chi/V_{1},
\delta \equiv V_{2}/V_{1}$)
\vspace{0.4cm}

\noindent{\bf Fig.3 :}\ \ Bound state energy as a function of positive (a) and negative(b) nonlinearity,
for several dispersion values.
\vspace{0.4cm}

\noindent{\bf Fig.4 :}\ \ Probability at the impurity site as a function of nonlinearity, in the negative
nonlinearity sector, for several dispersion values.
\vspace{0.4cm}

\noindent{\bf Fig.5:}\ \ Transmission coefficient of plane waves across the nonlinear impurity as a function
of the plane wave dimensionless energy, for several dispersion values.
\vspace{0.4cm}

\noindent{\bf Fig. 6:}\ \ Long-time average probability of finding the electron on the impurity site, as
a function of nonlinearity, for several different dispersion values. ($T = 203\ V_{1}$)
\vspace{0.4cm}

\noindent{\bf Fig. A:}\ \ Real part of the diagonal unperturbed Green function $G_{0 0}^{(0)}(z)$, for several
dispersion values:\ 0 (a), 0.2(b), 0.4(c), 0.6 (d), 0.8 (e) and 1.0(e).
\vspace{0.4cm}

\noindent{\bf Fig. B:}\ \ Same as in Fig. A, but for the imaginary part of $G_{0 0}^{(0)}(z)$.


\begin{thebibliography}{9}

\bibitem{economou1}
E.N. Economou and C.M. Soukoulis, Phys. Rev. B {\bf 28}, 1093 (1983);
E. N. Economou, C.M. Soukoulis and A.D. Zdetsis, {\em ibid.} {\bf 30},
1686 (1984).   

\bibitem{cmt}
See, for instance, D. Chen, M. I. Molina and G. P. Tsironis, J. Phys.: Condens. Matter {\bf 5},
8689 (1993); D. Chen, M. I. Molina and G. P. Tsironis, J. Phys.: Condens. Matter {\bf 8}, 6917 (1996).

\bibitem{tdm}
See, for instance, G. P. Tsironis, W. D. Deering and M. I. Molina,
Physica D {\bf 68}, 135 (1993) and references therein.

\bibitem{mt_prb}
M. I. Molina and G. P. Tsironis, Phys. Rev. B {\bf 47}, 15330 (1993);

\bibitem{mth_pre}
G. P. Tsironis, M. I. Molina and D. Hennig, Phys. Rev. E {\bf 50}, 2365 (1994).

\bibitem{saopaulo}
M.I. Molina in {\em Topics in Theoretical Physics}, edited by V.C.
Aguilera-Navarro, D. Galletti, B.M. Pimentel and L. Tomio, (IFT,
Sao Paulo, 1996); M.I. Molina, Phys. Rev. B {\bf 60}, 2276 (1999).

\bibitem{square}
M.I. Molina, Phys. Rev. B {\bf 60}, 2276 (1999);
K. M. Ng, Y.Y. Yiu  and P.M. Hui, Solid State Commun. {\bf 95}, 801 (1995).

\bibitem{cayley}
B. C. Gupta and S. B. Lee , Phys. Rev. B {\bf 63}, 144302 (2001).

\bibitem{hui}
Y. Y, Yiu, K. M. Ng and P. M. Hui, Phys. Lett. A {\bf 200}, 325 (1995)

\bibitem{busta}
C. A. Bustamante and M. I. Molina, Phys. Rev. B {\bf 62}, 15287 (2000).

\bibitem{gaididei}
Yu. B. Gaididei, S. F. Mingaleev, P. L. Christiansen and K. \O. Rasmussen,
Phys. Lett. A {\bf 222}, 152 (1996).

\bibitem{rasmussen}
K. \O. Rasmussen, P. L. Christiansen, M. Johansson, Yu. B. Gaididei
and S. F. Mingaleev, Physica D {\bf 113}, 134 (1998).

\bibitem{mt_prl}
M.I. Molina and G.P. Tsironis, Phys. Rev. Lett. {\bf 73},
464 (1994).

\bibitem{economou2}
E.N. Economou, {\em Green's Functions in Quantum Physics}, Springer
Series in Solid State Physics, Vol.7 (Springer-Verlag, Berlin, 1979).

\bibitem{kivshar}
W. Kr\'{o}likowski, U. Trutschel, M. Cronin-Golomb and C. Schmidt-Hattenberger,
Opt. Lett. {\bf 19}, 320 (1994); W. Kr\'{o}likowski and Y. Kivshar,
J. Opt. Soc. Am. {\bf 13}, 876 (1996).


\end{thebibliography}
\end{document}